\documentclass[final
  ]
  {aipproc}

\usepackage{graphicx}
\usepackage{epsfig}
\usepackage{mathptm}
\layoutstyle{6x9}

\let\cal\mathcal

\renewcommand{\theequation}{\arabic{equation}}

\newcommand{\pvalt}{\raise0.15ex\hbox{-}\mkern-11.5mu\int}
\newcommand{\bea}{\begin{eqnarray}}
\newcommand{\eea}{\end{eqnarray}}
\newcommand{\ben}{\begin{enumerate}}
\newcommand{\een}{\end{enumerate}}
\newcommand{\bit}{\begin{itemize}}
\newcommand{\eit}{\end{itemize}}

\newcommand{\lab}[1]{\label{#1}}

\newcommand{\Eqs}[2]{Eqs.~(\ref{#1}) and (\ref{#2})}

\newcommand{\half}{\frac{1}{2}}

\newcommand{\p}{\partial}

\renewcommand{\ln}{\,\mbox{ln}\,}

\newcommand{\eps}{\epsilon}

\newcommand{\la}{\lambda}

\newcommand{\si}{\sigma}

\newcommand{\beq}{\begin{equation}}
\newcommand{\eeq}{\end{equation}}
\newcommand{\ba}{\begin{array}}
\newcommand{\ea}{\end{array}}

\begin{document}

\title{Unnatural Acts:\\ Unphysical Consequences of Imposing Boundary 
Conditions on Quantum Fields}

\author{ROBERT L. JAFFE}{
  address={Center for Theoretical Physics\\
Laboratory for
Nuclear Science and Department of Physics\\ 
Massachusetts Institute of Technology\\
Cambridge, MA 02139\\
E-mail: jaffe@mit.edu}
}



\begin{abstract}
I examine the effect of trying to impose a Dirichlet boundary
condition on a scalar field by coupling it to a static background. 
The zero point -- or Casimir -- energy of the field diverges in the
limit that the background forces the field to vanish.  This divergence
cannot be absorbed into a renormalization of the parameters of the
theory.  As a result, the Casimir energy of a surface on which a
Dirichlet boundary condition is imposed, and other quantities like the
surface tension, which are obtained by deforming the surface, depend
on the physical cutoffs that characterize the coupling between the
field and the matter on the surface.  In contrast, the energy density
away from the surface and forces between rigid surfaces are finite and
independent of these complications.
\end{abstract}
\begin{flushright}CTP-MIT-3394\end{flushright}
\maketitle


\section{Introduction}

This talk is offered to Joe Schechter on the occasion of his 65th
birthday.  Joe has always looked carefully at things that others took
for granted, and often discovered new and interesting physics as a
result.  Here I take a look at what boundary conditions do to the high
energy behavior of interacting fields, and find some unexpected
results.  I hope Joe finds the story interesting and the argument
convincing!  The work described here is a by-product of a large
collaboration including Eddie Farhi, Noah Graham, Vishesh Khemani,
Markus Quandt, Marco Scandurra, Oliver Schr\"oder, and Herbert Weigel. 
A summary of these results can be found in Ref.~\cite{Graham:2002fw}
and in a forthcoming publication.~\cite{intheworks}

Obviously boundary conditions are a very convenient idealization in
field theory.  Conducting boundary conditions give an excellent
description of the behavior of electric and magnetic fields near good
metals.  Bag boundary conditions give a very useful characterization
of the low modes of the quark field in hadrons.  However physical
materials cannot constrain arbitrarily high frequency components of a
fluctuating quantum field, so the use of boundary conditions to model
the effect of vacuum fluctuations, for example in the study of the
Casimir effect, requires careful examination.  Boundary conditions
also appear in brane world scenarios for physics beyond the standard
model and in lattice implementations of quantum field theories.

Our\cite{Graham:2002fw,intheworks} point of view is that interactions
between fields and matter are fundamental and that boundary conditions
can only be substituted when they can be shown to yield the same
physics.  Surprisingly, there are important cases where they do not. 
The issue is the appearance of divergences or cutoff dependence in
calculations of the total quantum zero point energy (the ``Casimir
energy'') of a field subject to a boundary condition.  If boundary
conditions are placed on the field {\it ab initio\/}, methods have
been developed that appear to yield a finite, cutoff independent
energy.\cite{MT} In contrast, we write down a quantum field theory
describing the interaction of the fluctuating field with a static
background, representing matter, and take a limit involving the
shape of the background and the strength of the interaction that
produces the desired boundary condition on a specified surface.  Since
the initial field theory is renormalizable, a finite, renormalized
Casimir energy can be defined and computed for any non-singular
background.  The renormalized Casimir energy diverges in 
the boundary condition limit, indicating that the physical vacuum 
fluctuation energy depends in detail on the properties of the material 
that provides the physical ultraviolet cutoff.  Note that these 
divergences have nothing to do with the standard infinities of quantum 
field theory, which were removed by renormalization.  Instead they 
arise because the background necessary to enforce the boundary 
condition has too much strength at high frequencies.

I don't want the point to get lost in complicated algebra, so I will
discuss a case so simple that the necessary calculations are
elementary.  I will consider a scalar field in one dimension, obeying
the Dirichlet boundary condition, $\phi=0$, at one or two points.  The
limitation of this simple example is that the novel effects  
show up in the Casimir energy, but not in the Casimir force.  In higher
dimensions the effects are more dramatic and do effect measureable
quantities.  However the calculations are harder, so I will only
report our results which are described in detail
elsewhere.\cite{Graham:2002fw,intheworks} Of course this subject has
been treated before, but not, as far as I know, in quite the way that
I will describe.  I will mention some of the other treatments later in
my talk.

\section{Dirichlet Points -- A Toy Model in One Dimension}

As a warm up, consider what it takes to enforce the boundary
condition, $\phi(0)=0$ on an otherwise non-interacting and massless
scalar quantum field in one dimension.  This is the problem of the
``Dirichlet point''.  The equation of motion for $\phi$ coupled to
some otherwise inert potential, $\sigma(x)$, is
$$
-\phi'' +\lambda\sigma(x)\phi+m^{2}\phi=\omega^{2}\phi
$$ 
for a mode with energy $\omega$.  The coupling $\lambda$ is defined by 
normalizing $\sigma$ so that $\int\sigma(x)=1$.  

To get all eigenmodes of $\phi$ to vanish at $x=0$ it is necessary to
take $\sigma$ to be ``sharp'' and $\lambda$ to be ``strong'':
\begin{eqnarray*}
	\sigma(x)&\to\delta(x) \quad\ \hbox{\sc sharp}\\
	\lambda & \to\infty \quad \ \ \ \ \ \hbox{\sc strong}
\end{eqnarray*}
In the sharp limit $\phi'$ suffers a discontinuity at $x=0$,
$\Delta\phi'|_{0}=\lambda\phi(0)$.  A careful study of the limit
$\la\to\infty$ shows that the boundary condition $\phi(0)=0$ emerges
for all $\omega$.  {\it A priori\/} one would expect such a strong
interaction to have a dramatic effect on the dynamics of $\phi$, in
particular on sum over zero point energies with the boundary condition
compared to the energy without.

\begin{figure}[th]
\centerline{
\includegraphics{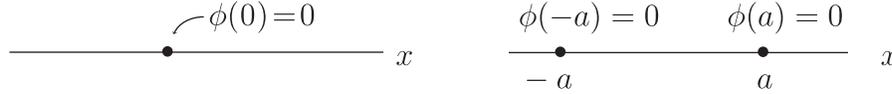}  
}
\caption{Dirichlet points: a) one, and b) two at a separation of $2a$}
\label{dirichletpoints}
\end{figure}

First, suppose the boundary condition is imposed at the outset on all
modes.  This is the standard approach.\cite{MT} The vacuum
fluctuation energy is the sum over zero point energies of $\phi$ minus
the sum without the boundary condition,
\begin{equation}
	\tilde E_{1}=\half\sum \left(\hbar\omega-\hbar\omega_{0}\right)
\end{equation}
The situation is shown in Fig.  \ref{dirichletpoints}a.
The solutions to the free field theory are $\phi\sim\sin kx$ and
$\phi\sim\cos kx$ for $k>0$.  With the boundary condition $\phi(0)=0$
the solutions are $\phi\sim\sin kx$ and $\phi\sim\sin|k|x$, also with
$k>0$.  Since the spectra are identical, $\omega$ and $\omega_{0}$
simply cancel and leave
\begin{equation}
	\tilde E_{1}=0
	\lab{we1}
\end{equation}
The tilde is there to remind us that the boundary condition was
imposed {\it ab initio\/}.  The treatment in the standard texts is more
careful than this, but in essence the same.  Next, apply the same
methods to the case of two Dirichlet points located at $x=\pm a$ (see
Fig.~\ref{dirichletpoints}b).  The textbook result is
\begin{equation}
	\tilde E_{2}(a)=-\frac{\pi}{48a}\quad\hbox{for}\quad m=0
	\lab{we2}
\end{equation}

\Eqs{we1}{we2} are actually bizarre and unacceptible.  First, remember
that these are the \emph{total} energies for each configuration
relative to the vacuum -- \emph{we did not drop any terms}.  To see the
problem, consider first the limit $a\to\infty$:
\begin{equation}
	\lim_{a\to\infty}\tilde E_{2}(a)=0=2\tilde E_{1}
\end{equation}
which is fine.  It confirms that two widely separated Dirichlet points 
do not interact.  Now consider the limit $a\to 0$,
\begin{equation}
	\lim_{a\to 0}\tilde E_{2}=-\infty \stackrel{?}{=}\tilde E_{1}
	\lab{wrlimit}
\end{equation}
which seems to say that the energy of a single Dirichlet point, the 
result of two that coalesce as $a\to 0$, is infinite.  Lastly, note 
that all of these results are suspicious because the massless scalar 
field theory suffers from infrared divergences in one dimension, 
leading us to expect log divergences where none have been found.

Now let us consider the same problem from the perspective of an
interaction of $\phi$ with matter.  We make the minimal model: we
couple $\phi$ to a \emph{non-dynamical} scalar background field,
$\sigma(x)$, with the (super) renormalizable interaction,
\begin{equation}
	{\cal L}  = \half\p_{\mu}\phi\p^{\mu}\phi -\half m^{2}\phi^{2}
	-\lambda\sigma(x)\phi^{2} + c(\epsilon)\sigma(x)
	\lab{lint}
\end{equation}
Where $c(\eps)\si(x)$ is a counterterm and $\eps$ is a cutoff, for
example the fractional part of the dimension in dimensional
regularization.  We could elevate $\sigma$ to be a dynamical field by
endowing it with an action of its own.  However the core of the
problem can be studied without this complication.  ${\cal L}$
describes a renormalizable quantum field theory.  In one dimension
only one Feynman diagram -- the tadpole -- is divergent.  This
divergence is cancelled by the counterterm.  As usual there is some
scheme dependence involved in renormalization.  We choose the ``no
tadpole scheme'' where $c(\epsilon)$ is chosen to completely cancel
the tadpole graph so that $\langle\sigma\rangle=0$.  Any other scheme
can be related to ours by a trivial shift in $\sigma$.  This theory
has infrared divergences as $m\to 0$, so we keep $m\ne 0$ throughout.

It is an straightforward to compute the renormalized energy of any
configuration, $E[\sigma]$, relative to the vacuum, $\sigma=0$, and to
show that it is finite for piecewise continuous $\sigma(x)$.  The
crucial question is what happens to $E[\sigma]$ when we try to take
$\sigma$ to be \emph{sharp} and $\lambda$ to be \emph{strong}.  Do we
reproduce the results obtained when the boundary condition is imposed
{\it ab initio\/} or not?
\begin{figure}[th]
\centerline{
\includegraphics{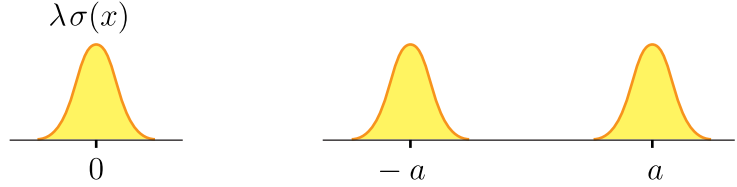}   
}
\caption{Smooth but sharply peaked backgrounds approximating Dirichlet 
points: a) one,
and b) two at a separation of $2a$.  In the ``sharp'' limit the 
backgrounds approach delta functions.}
\label{lumps}
\end{figure} 
The \emph{sharp} limit is benign in one dimension, so we can take
$\sigma(x)$ to be a ``spike'' $\sigma(x)=\delta(x)$ without
difficulty.  The renormalized energy of an isolated spike of strength
$\lambda$ (see Fig.~\ref{lumps}a) is found to be
\begin{equation}
	E_{1}(\lambda,m) = \frac{1}{2\pi}\int_{m}^{\infty}dt
	\frac{t\log\left(1+\frac{\lambda}{2t}\right)-\frac{\lambda}{2}}
	{\sqrt{t^{2}-m^{2}}}
	\lab{re1}
\end{equation}
It is easy to see that this integral converges for any $\lambda$ and
$m\ne 0$, but diverges logarithmically as $m\to 0$.  Likewise, the
renormalized energy of two spikes at $x=\pm a$ is given by,
\begin{equation}
	   E_2(a,\lambda,m) = \frac1{2\pi} \int_m^\infty\!  dt\,
	   \frac{t\log \left(1+\frac\lambda{t} + \frac{\lambda^2}{4t^2}
	   (1-e^{-4at})\right) - \lambda}{\sqrt{t^2-m^2}}   
	   \lab{re2}
\end{equation}
Since these results are central to the rest of the discussion, I've 
appended a derivation (of eq.~(\ref{re1})) at the end of the talk.

Let's submit these results to the same examination as before:  as 
$a\to\infty$ we find
\begin{equation}
	\lim_{a\to\infty}E_{2}(a,\lambda,m) = 2E_{1}(\lambda,m)
\end{equation}
which is fine; and as $a\to 0$,
\begin{equation}
	\lim_{a\to 0}E_{2}(a,\lambda,m)=E_{1}(2\lambda,m)
\end{equation}
which is much more satifactory than eq.~(\ref{wrlimit}). Also, both 
$E_{1}$ and $E_{2}$ diverge logarithmically as $m\to 0$, as expected.  
So the renormalized energy in a sharp background is finite and passes 
all the tests.

However, the punchline is that \emph{both $E_{1}$ and $E_{2}$ diverge
like $-\lambda\ln\lambda$ as $\lambda\to\infty$}.  So the total
\emph{renormalized} Casimir energy diverges as the coupling constant,
$\lambda$, is taken strong enough to impose the boundary condition on
all modes.  $\lambda$ can be regarded as a cutoff, since modes with
$\omega\gg\lambda$ are not affected by the interaction.  So in this
example in a physical material the total, renormalized vacuum
fluctuation energy is strongly cutoff dependent.  Note that $E_{1}$
and $E_{2}$ are \emph{renormalized} energies.  There are no
counterterms available to absorb this divergence.
	
The two approaches disagree on the total energy, but they agree on the
energy density and the force between two points.  For any background
$\sigma(x)$, the energy density diverges only where $\sigma(x)\ne 0$. 
For a spike at $a$ the energy density remains finite for $x\ne \pm a$
in the limit $\lambda\to\infty$, and the limiting form agrees with the
density calculated using the boundary condition {\it a
priori}.\cite{MT} Likewise, the force between the two sharp sources
agrees with the boundary condition calculation as $\lambda\to\infty$,
\begin{equation}
	\lim_{\lambda\to\infty}-\frac{\p}{\p a}E_{2}(a,\lambda,m)
	=-\frac{\p}{\p a} \tilde E_{2}(a,m)\ \left(\ = -\frac{\pi}{48a^{2}}
	\quad\hbox{for}\quad m=0\right)
\end{equation}
In fact no measurement of the properties of the two points can detect
the infinite energy stored locally on the boundary.  This, however, is
special to one dimension.  Notice also that the vacuum fluctuation
energy is  negative ($\sim-\lambda\ln\lambda$ as
$\lambda\to\infty$).  In a more realistic context this energy would be
more than overwhelmed by the positive contributions to the energy
coming from the curvature of $\sigma$, which goes like $|\sigma'|^{2}$,
and the potential energy of $\sigma(x)$ which would involve higher
powers of $\sigma(x)$ beginning with the mass term $\half m^{2}\int dx
\sigma^{2}(x)$.  I left those terms out, not because they
aren't present, but because I was trying to define an abstract problem
-- the ``Dirichlet-Casimir'' problem.  Having failed, it is clear that
the total energy depends not only on the vacuum fluctuation energy of
$\phi$, but also on the energy stored in the material represented by
$\sigma$.  The two cannot be separated -- no abstract ``Casimir energy'' 
can be defined for the Dirichlet boundary condition in one dimension.

To summarize the results for a scalar field in one dimension:
\begin{itemize}
	\item The renormalized vacuum fluctuation energy is well defined
	and finite for any (piecewise continuous) background $\sigma(x)$. 
	It differs from the energy calculated by assuming a boundary
	condition {\it ab initio\/}.  The renormalized energy is material
	({\it ie.\/} $\lambda$), dependent, and diverges if $\lambda$ is
	taken to infinity to impose the boundary condition on all modes.

	\item The change in the vacuum fluctuation energy with rigid 
	displacement of the boundaries is finite and 
	cutoff independent and can be calculated by imposing the boundary 
	condition {\it ab initio\/}.
\end{itemize}
 
\section{Physical Effects in Other Dimensions}

The result of the previous section would be a merely a curiosity if it
could not be measured.  In higher dimensions worse divergences occur
and they too are always confined to the domain where the background fields
are non-zero.  Thus in any experiment in which the material ($\equiv$
the background fields) are unchanged, the cutoff dependence will
cancel out.  A case in point is the standard Casimir force between
parallel, grounded, conducting plates.  The force is measured by
displacing the plates rigidly but not deforming them.  The cutoff
dependent terms remain unchanged as the plates are moved and the 
resulting force is finite.  The result is the same whether you impose 
the boundary condition at the beginning or start with a smooth 
background and impose the boundary condition in a limiting process.

The situation is completely different, however, if the material must
be deformed to display the physical effect.  The ``Casimir pressure''
on a sphere is the most interesting example .  To measure this it is
necessary to compare the total Casimir energy of a sphere of radius
$R$ with that of a sphere of radius $R+\delta R$.  If there are cutoff
dependent terms associated with the material, their contribution to
the energy changes as the area changes, and they contribute to the
pressure.  Thus, calculations of Casimir pressures will differ between
the case where boundary conditions are applied ${\it ab\ initio\/}$ and
where they are realized as a limit of the coupling to a background
field.

To examine these issues quantitatively we have studied the problem of
the ``Dirichlet sphere'' -- the boundary condition $\phi(R)=0$ imposed
on a massive (or massless) scalar field in $D$
dimensions.  \footnote{Here I am describing work reported briefly in
Ref.~\cite{Graham:2002fw}.  A more complete discussion is nearing 
completion in Ref.~\cite{intheworks}} $D=2$ is the ``Dirichlet
circle'' and $D=3$ is the sphere.  As in the one dimensional example,
we replace the boundary condition by the coupling to a non-dynamical
scalar background, $\sigma$, with an interaction of the form 
$$
{\cal
L}_{\rm int}=\lambda \sigma(r)\phi^{2}+c_{1}(\epsilon)\sigma +c_{2}(\epsilon)
\sigma^{2}+ \ldots,
$$
where the $\ldots$ denote a finite series of counterterms necessary to
renormalized the perturbative divergences generated by the
$\phi-\sigma$ interaction.  For $D=2$ and $D=3$ only the counterterms
shown are needed.   We normalized the source so its integral over space is 
one,
$$
\int d^{D}r \si(r)=1
$$
so the coupling constant, $\la$, has dimension $[{\rm
mass}]^{2-D}$\footnote{This corresponds to the physical situation
that the surface gets thinner as $R$ is increased.}

\begin{figure}[th]
\centerline{
\includegraphics[width=5.75in]{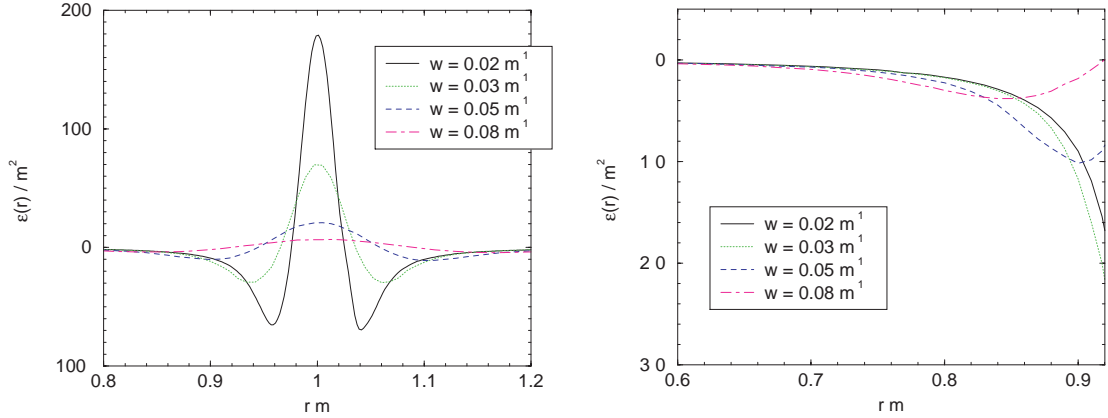}
}
\caption{Casimir energy density in a Gau{\ss}ian background of width 
$w$ approximating a Dirichlet circle.  a) The energy density; b) a 
closeup of the behavior near $r=0.8/m$.}
\label{dirichletcircle}
\end{figure}

We then compute the one-loop effective energy ({\it ie.\/} the Casimir
energy), $E_{D}(R,\Delta, \lambda)$, for a background $\sigma$ which
is peaked at $r=R$ and has a thickness $\Delta$.  We have studied
Gau{\ss}ians and square barriers.  The renormalized Casimir energy is
finite for fixed $\Delta$, and $\lambda$.  However it diverges in the
sharp limit, $\Delta\to 0$ (except in one dimension, where we saw that
the sharp limit was benign), so that not even the sharp limit exists
for $D\ge 2$.  If we keep $\Delta$ fixed, and take $\lambda\to\infty$
the Casimir energy diverges as well.  The nature of the divergences
depends on $D$.
\begin{itemize}
	\item For the Dirichlet circle ($D=2$), $E_{2}$ diverges like
	$\frac{\lambda^{2}}{R}\log\Delta $ at fixed $\lambda$ as
	$\Delta\to 0$.  Thus the tension (the two dimensional analog
	of pressure) diverges logarithmically as the thickness of the
	``circle'' goes to zero.
	
	\smallskip \noindent This is particularly nicely illustrated
	by examining the energy density, $\eps(r)$, in a spherical
	shell between $r$ and $r+dr$.  The necessary formalism was
	developed in Ref.~\cite{Graham:2002xq} In
	Fig.~\ref{dirichletcircle}a we plot $\eps(r)$ for a
	Gau{\ss}ian background of fixed strength as a function of its
	width (denoted $w$ in the figures).  As $w\to 0$ the energy
	density diverges in a non-uniform way: at any fixed $r\ne R$,
	it approaches a limit.  However the closer to $R$, the slower
	the convergence, and as a result, the total energy diverges as
	$w\to 0$.  The non-uniform behavior is clear in
	Fig.~\ref{dirichletcircle}b.
	
	\item For the Dirichlet sphere ($D=3$), the leading divergence in
	$E_{3}$ goes like $\frac{\lambda^{2}}{R^{2}}\frac{\log\Delta}{\Delta}$. 
	Thus the pressure diverges as the thickness of the sphere goes to 
	zero.
\end{itemize}

We expect that these divergences signal that the actual energy depends
on the physical properties of the material.  $\Delta$ represents its
thickness and $\lambda$ plays the role of a frequency cutoff ({\it
eg.\/} the plasma frequency, $\omega_{p}$) since modes with
$\omega\gg\lambda$ are unaffected by $\sigma$.  Since the other
stresses to which materials are subject also depend on such properties
of the material, we conclude that \emph{the Casimir pressure on a
surface cannot be defined in a useful way that is independent of the
particular dynamical properties of the material}.  

Since the Casimir energy for any piecewise continuous background,
$\sigma$, has an expansion in Feynman diagrams, and since the
divergence structure of Feynman diagrams is very well understood, it
is interesting to look for the origin of the divergences in the
diagrams.  The Casimir energy is simply the sum of all one-loop
diagrams for $\phi$ circulating in the background $\sigma$.  We find
that the divergences \emph{never} come from the loop integrations. 
Renormalization takes care of those.  Instead the divergences come
from the integration over the Fourier components of the external
lines, $\tilde \sigma (p)$.  For $D=2$ the divergence comes from the
renormalized two-point function.  Since the two-point function is easy
to calculate, the divergence is easy to display.  For $D=3$ the
two-point function also generates the leading divergence.  However
even the three-point function also diverges (logarithmically) for
$D=3$.  Since the loop integration for the three-point function is
manifestly convergent, it provides a very clear demonstration that
the divergences as $\Delta\to 0$ cannot be renormalized away by any
counterterm available in the continuum theory.  Instead they are
physical manifestations of the cutoff dependence of the Casimir energy
for the Dirichlet sphere.

Our result disagrees with calculations which assume a boundary 
condition  {\it ab initio\/}.\cite{Milton:2002vm}  However those 
calculations handle divergences in an {\it ad hoc\/} manner and obtain 
suspicious results:  for example the Casimir energy of a Dirichlet 
sphere is claimed to be finite in odd dimensions
($D=1,3,5,7 \ldots$) and to diverge in even dimensions.

The study of field theories in the presence of boundaries is an old
subject.  The divergences that occur near boundaries have been studied
by many authors and with powerful tools.\cite{Symanzik:1981wd} In
particular, it is well known that there are serious divergences near
boundaries.  Symanzik showed that all the divergences in the presense
of boundaries (in certain theories) can be systematically cancelled by
the introduction of new counterterms that live on the boundaries. 
However this begs our question of whether the divergences can be
cancelled with only the counterterms available in an underlying
interacting quantum field theory without boundaries.  As far as I 
know, the implementation of boundary conditions as a limit of 
renormalizable, interacting quantum field theories is new.

Although we have only completed our study for the Dirichlet boundary
condition, we have examined conducting boundary conditions as well and
expect our conclusions to apply to that case as well.
 
Our interest in vacuum fluctuation energies goes far beyond the 
question I have discussed today.  We are interested in whether quantum 
fluctuations can generate novel and complex behavior like solitons, in 
realistic quantum field theories.  However, our investigations led us 
to a close examination of the nature of Casimir effects and to a 
result surprising enough to be worth telling Joe on the occasion of 
his birthday!


\begin{theacknowledgments}
 This work has been done in collaboration with N.~Graham, V.~Khemani,
M.~Quandt, M.~Scandurra, O.~Schr\"oder, and H.~Weigel.  The work I
have reported is in large measure theirs.  My work is supported in
part by the U.S.~Department of Energy (D.O.E.) under cooperative
research agreement~\#DF-FC02-94ER40818.
\end{theacknowledgments}

\section*{Appendix}

Here is an outline of the calculation of the vacuum fluctuation energy 
of an isolated ``spike'', eq.~(\ref{re1}).  We start with the 
universally accepted formula for the sum over zero point energies of a 
field, $\phi$, fluctuating in a background $\sigma$:
\begin{equation}
E[\sigma]=\half\sum_{j} \omega_{j} +\frac{1}{2}
\int_{0}^{\infty}dk\sqrt{k^{2}+m^{2}}\ \delta\rho(k)
\lab{ecas}
\end{equation}
This is just the generalization to the continuum of
$\half\sum\hbar\omega-\hbar\omega_{0}$ (with $\hbar=1$ of course). 
The $\omega_{j}$ are the energies of possible bound states. 
$\delta\rho(k)$ is the change in the density of states due to the
background $\sigma$.  $\delta\rho$ is given by another well known
result:
$$
\delta\rho(k)=\frac{1}{\pi}\sum_{\ell}\frac{d\delta_{\ell}}{dk}\ ,
$$
where $\delta_{\ell}(k)$ is the phase shift for scattering in the 
$\ell^{\rm th}$ partial wave in the background $\sigma$ (assumed to be 
symmetric).  

$E[\sigma]$ defined by eq.~(\ref{ecas}) diverges in one dimension.  To
keep control of divergences we use dimensional regularization: we
imagine that we are computing in $D$ dimensions.  It can be shown that
eq.~(\ref{ecas}) is finite for $0<D<1$.  After isolating the possible
divergences and renormalizing we will analytically continue to $D=1$. 

Next we use Levinson's theorem -- which equates $\pi$ times the number
of bound states to the difference of the phase shift at $k=0$ and
$k=\infty$ (valid in $D$ dimensions) to rewrite $E[\sigma]$ as
\begin{equation}
E[\sigma]=\half\sum_{j} (\omega_{j}-m) +\frac{1}{2\pi}
\int_{0}^{\infty}dk(\sqrt{k^{2}+m^{2}}-m)\sum_{\ell}\frac{d\delta_{\ell}}{dk}
\lab{ecasr}
\end{equation}
At this point it is possible to show that the successive terms in the 
Born expansion of $\delta_{\ell}$ (the expansion of $\delta$ in powers 
of $\sigma$) can be put into one-to-one correspondence with the one 
loop Feynman diagrams.\footnote{This is not true until after the 
Levinson's subtraction has been made.  For a discussion and references, 
see \cite{Graham:2002fi}.} In particular the first Born approximation 
is identical to the tadpole diagram which diverges as $D\to 1$.  

To renormalize, we first subtract the first Born approximation from 
eq.~(\ref{ecasr}) and add back the tadpole diagram to which it is equal.  
We then cancel the tadpole diagram against the counterterm 
$c(\epsilon)$ (see eq.~(\ref{lint})).  The resulting 
\emph{renormalized} and manifestly finite expression for the Casimir 
energy can now be evaluated at $D=1$, where the sum over phase shifts 
includes only the symmetric and antisymmetric channels:
\begin{equation}
E[\sigma]= \frac{1}{2\pi}
\int_{0}^{\infty}dk(\sqrt{k^{2}+m^{2}}-m) \frac{d}{dk}\left(
\delta_{+}(k)+\delta_{-}(k)-\delta^{(1)}(k)\right)
\lab{ecast}
\end{equation}
$\delta^{(1)}(k)$ is the Born approximation to the sum of the 
phase shifts.  To save space I have dropped the sum over bound states 
since there are none in the case at hand.

Now to the specific case of the $\delta$-function background:  The
phase shift in the antisymmetric channel vanishes.  The symmetric
channel phase shift is easily computed:
$$
\delta_{+}(k)=\tan^{-1}\frac{\lambda}{2k}
$$ 
and the first Born approximation is
$$
\delta^{(1)}(k)=\frac{\lambda}{2k}
$$
The resulting integral,
$$
E(\lambda,m) = \frac{\lambda^{3}}{4\pi}\int_{0}^{\infty}dk
\frac{\sqrt{k^{2}+m^{2}}-m}{k^{2}(\lambda^{2}+4k^{2})}
$$ 
is most easily evaluated by rotating the contour to the positive 
imaginary axis and integrating by parts.  The result is the expression 
quoted in eq.~(\ref{re1}).




\end{document}